\documentclass[pra,aps,amsmath,amssymb,showkeys,eqsecnum]{revtex4}
\newcommand{\hh}{{\mathcal{H}}}

\newcommand{\pen}{\openone}

\newcommand{\tr}{\mathrm{tr}}

\newcommand{\rmc}{{\mathrm{C}}}
\newcommand{\rmh}{{\mathrm{H}}}
\newcommand{\bro}{\boldsymbol{\rho}}
\newcommand{\vbro}{\boldsymbol{\varrho}}
\newcommand{\bsg}{\boldsymbol{\sigma}}
\newcommand{\bau}{\boldsymbol{\tau}}

\newcommand{\clb}{{\mathcal{B}}}
\newcommand{\cle}{{\mathcal{E}}}
\newcommand{\clf}{{\mathcal{F}}}
\newcommand{\cli}{{\mathcal{I}}}

\newcommand{\cln}{{\mathcal{N}}}

\newcommand{\mm}{{\mathsf{M}}}

\newcommand{\me}{{\mathsf{E}}}
\newcommand{\km}{{\mathsf{K}}}
\newcommand{\sm}{{\mathsf{S}}}
\newcommand{\qm}{{\mathsf{Q}}}
\newcommand{\um}{{\mathsf{U}}}
\newcommand{\matph}{{\pmb{\Phi}}}
\newcommand{\mtgm}{{\pmb{\Upsilon}}}
\newcommand{\iu}{{\mathtt{i}}}

\newcommand{\vark}{\varkappa}

\begin{document}
\clearpage
\preprint{}

\title{Entropic uncertainty relations from equiangular tight frames and their applications}

\author{Alexey E. Rastegin}
\affiliation{Department of Theoretical Physics, Irkutsk State
University, K. Marx St. 1, Irkutsk 664003, Russia}

\begin{abstract}
Finite tight frames are interesting in various topics including
questions of quantum information. Each complex tight frame leads to
a resolution of the identity in the Hilbert space. Symmetric
informationally complete measurements are a special class of
equiangular tight frames. Applications of such frames in quantum
physics deserve more attention than they have obtained. We derive
uncertainty relations for a quantum measurement assigned to an
equiangular tight frame. Main results follow from estimation of the
corresponding index of coincidence. State-dependent and
state-independent formulations are both addressed. Also, we discuss
applications of considered measurements to detect entanglement and
other correlations.
\end{abstract}

\keywords{finite frames, generalized entropies, uncertainty relations, entanglement detection}

\maketitle

\pagenumbering{arabic}
\setcounter{page}{1}

\section{Introduction}\label{sec1}

Studies of discrete structures in finite-dimensional Hilbert spaces
has a long history \cite{lint66,seidel73}. Such structures are
interesting not only in their own rights but also due to potential
application in quantum physics. Emerging technologies of quantum
information processing gave a new stimulus to investigate finite
sets of states with special properties. Mutually unbiased bases
(MUBs) are one of the most known examples of this kind
\cite{ivano81,fields89}. Another remarkable paradigm of discrete
structures is given by symmetric informationally complete
measurements \cite{zauner11,rbksc04}. Quantum and unitary designs
are considered as a powerful tool of quantum information theory
\cite{gross07,cirac18,dcel9,chart19}. On the other hand, the
question of building such structures are often difficult to resolve
\cite{fhs2017,abfg19}. For instance, the maximal number of MUBs
remains unknown even for $d=6$, i.e., for the smallest dimension
that is not a prime power.

Finite tight frames are a natural generalization of orthonormal
bases \cite{waldron18}. Each of such frames can be applied to build
a positive operator-valued measure (POVM). POVMs are an
indispensable tool in quantum information science \cite{nielsen}.
Tight frames have also found use in signal processing and coding
\cite{caskut}. In equiangular tight frames (ETFs), any two frame
vectors have the same overlap. In a certain sense, this idea is
similar to that is used to define MUBs. Maximal sets of complex
equiangular vectors provide symmetric informationally complete POVMs
\cite{waldron18}. The existence of such POVMs for arbitrary
dimensions is still an open question, though several exact
constructions have been found \cite{fhs2017,abb19,yard2020}. This
gives a reason to study arbitrary ETFs in more detail and use them
together with the maximal ones.

In this paper, we consider entropic uncertainty relations for a
quantum measurement built of the states of an ETF. Potential
applications in quantum information science will be mentioned as
well. The paper is organized as follows. In Sec. \ref{sec2}, the
required material on complex tight frames is briefly recalled.
Section \ref{sec3} is devoted to estimate from above the
corresponding index of coincidence, whence many results will be
obtained. Fine-grained uncertainty relations are examined in Sec.
\ref{sec4}. In Sec. \ref{sec5}, we formulate entropic uncertainty
relations for quantum measurements assigned to ETFs. To quantify the
amount of related uncertainties, the R\'{e}nyi and Tsallis entropies
are both used. In Sec. \ref{sec6}, the derived relations are applied
to entanglement detection, inequalities for conditional von Neumann
entropies and quantum coherence. In Sec. \ref{sec7}, we conclude the
paper with a summary of the results.

\section{Preliminaries}\label{sec2}

In this section, we recall some material concerning ETFs.
The authors of \cite{sustik,mixon} discussed the
existence of such frames in both the real and complex cases. In the
following, all the frames are assumed to be complex. Let $\hh_{d}$
be a $d$-dimensional Hilbert space. A set of $n\geq{d}$ unit vectors
$\clf=\bigl\{|\phi_{j}\rangle\bigr\}$ is called a frame if there
exist strictly positive numbers $S_{0}<S_{1}<\infty$ such that
\begin{equation}
S_{0}\leq\sum\nolimits_{j=1}^{n}\bigl|\langle\phi_{j}|\psi\rangle\bigr|^{2}
\leq{S}_{1}
\label{abframe}
\end{equation}
for all unit $|\psi\rangle\in\hh_{d}$. The numbers $S_{0}$ and
$S_{1}$ are the minimal and maximal eigenvalues of the frame
operator
\begin{equation}
\sm=\sum\nolimits_{j=1}^{n}|\phi_{j}\rangle\langle\phi_{j}|
\, . \label{sframe}
\end{equation}
The special case $S_{0}=S_{1}=n/d$ gives a tight frame. Then the
frame operator is scalar with the eigenvalue $S=n/d$ of multiplicity
$d$. Parseval tight frames obtained with $S=1$ are equivalent to
orthonormal bases. A special kind of tight frames is known as
equiangular ones. The tight frame $\clf$ is called
equiangular, when there exists $c>0$ such that
\begin{equation}
\bigl|\langle\phi_{i}|\phi_{j}\rangle\bigr|^{2}=c
\label{md2c}
\end{equation}
for each pair $i\neq{j}$. By calculations, for an ETF, we have
\begin{equation}
S=nc+1-c=\frac{n}{d}
\ , \qquad
c=\frac{S-1}{n-1}=\frac{n-d}{(n-1)d}
\ . \label{abeq}
\end{equation}
If there is an ETF with $n$ elements in dimension $d$, then
$n\leq{d}^{2}$ and also exists an ETF with $n$ elements in dimension
$n-d$ \cite{sustik,mixon}. The second fact deserves to be considered
in more detail. Let us put the $d\times{n}$ matrix
\begin{equation}
\matph=\bigl(|\phi_{1}\rangle\,\cdots\,|\phi_{n}\rangle\bigr)
\, , \label{matm}
\end{equation}
where the frame states stand as columns. Since
$(d/n)\matph\matph^{\dagger}=\pen_{d}$ for an ETF,
rescaled rows of $\matph$ form an orthonormal set. One can convert
$(d/n)\matph\matph^{\dagger}$ into a unitary $n\times{n}$ matrix by
adding $n-d$ rows that are mutually orthogonal as well. Collecting
these rows into $(n-d)\times{n}$ matrix and normalizing its columns
gives other ETF \cite{mixon}. The latter contains $n$ vectors in
dimension $n-d$. This way is very close to the method used to
build a Naimark extension of rank-one POVM with $n$ elements. It is
important here that the given approach is constructive. In the
least case $n={d}^{2}$, it is seen from (\ref{md2c}) and (\ref{abeq})
that
\begin{equation}
\bigl|\langle\phi_{i}|\phi_{j}\rangle\bigr|^{2}=\frac{1}{d+1}
\qquad (i\neq{j})
\, . \nonumber
\end{equation}
Here, we deal with a symmetric informationally complete measurement
(SIC-POVM). As was already mentioned, there is a way to generate new
ETFs from the given ones. In particular, a $d$-dimensional
SIC-POVM induces ETFs in dimensions $d(d+1)/2$ and $d(d-1)/2$
\cite{abfg19}. Recently, some generalizations of the above ideas
were proposed. The authors of \cite{padg21} introduced the concept
of mutually unbiased frames as a general framework for studying
unbiasedness. Equioverlapping measurements were considered in
\cite{feng22}.

The states of an ETF induce the resolution $\cle=\{\me_{j}\}$ of
the identity, namely
\begin{equation}
\frac{d}{n}\>\sm=\sum\nolimits_{j=1}^{n} \me_{j}=\pen_{d}
\, . \label{sres}
\end{equation}
Here, the POVM elements are expressed as
\begin{equation}
\me_{j}=\frac{d}{n}\>
|\phi_{j}\rangle\langle\phi_{j}|
\, . \label{sres1}
\end{equation}
When the pre-measurement state is described by density matrix $\bro$
with $\tr(\bro)=1$, the probability of $j$-th outcome is equal to
\begin{equation}
p_{j}(\cle;\bro)=\frac{d}{n}\,\langle\phi_{j}|\bro|\phi_{j}\rangle
\, . \label{prbj}
\end{equation}
Developing some ideas of \cite{rastmubs}, we aim to derive entropic
uncertainty relations for POVMs assigned to ETFs. The composed
technique is similar to the method used for mutually unbiased bases
\cite{molm09}. Properties of the mentioned discrete structures allow
one to impose certain restrictions on generated probabilities. The
authors of \cite{gour2014} introduced the notion of general
symmetric informationally complete measurements. These measurements
are not necessarily described by elements of rank one. Similarly,
the set of $d+1$ mutually unbiased measurements can be built for
arbitrary $d$, when rank-one projectors are not required
\cite{kalev2014}. Entropic uncertainty relations were obtained for
mutually unbiased measurements \cite{rastosid,feiqip} and general
SIC-POVMs \cite{rastsic}. However, excessive costs may be required
to implement measurements with elements that are not of rank one.
Thus, ETF-based measurements are still of interest. Due to the
results of \cite{davies}, we can often restrict a consideration to
POVMs with rank-one elements. For such measurements, there is a
Naimark extension with only $n-d$ extra dimensions (see, e.g., the
comments right after (\ref{matm})). In more detail, the question of
building a Naimark extension is addressed in section 9-6 of
\cite{peres}.

\section{On the index of coincidence}\label{sec3}

The index of coincidence is used in various questions of information
theory \cite{harr2001}. In cryptography, for example, it gives a
measure of the relative frequency of symbols in a ciphertext sample
\cite{mvov97}. Now, we aim to estimate this quantity defined as
\begin{equation}
I(\cle;\bro)=\sum_{j=1}^{n} p_{j}(\cle;\bro)^{2}
\, . \label{indef}
\end{equation}
To each $|\psi\rangle\in\hh_{d}$, one assigns
$|\psi^{*}\rangle\in\hh_{d}$ as the ket with the conjugate
components in the canonical basis, whence
$\langle\varphi^{*}|\psi^{*}\rangle=\langle\varphi|\psi\rangle^{*}=\langle\psi|\varphi\rangle$.
Let us begin with an auxiliary result.

\newtheorem{aropn1}{Proposition}[section]
\begin{aropn1}\label{pnn1}
Let $n$ unit kets $|\phi_{j}\rangle$ form an equiangular tight frame
in $\hh_{d}$, and let
\begin{align}
|\Psi_{0}\rangle
&=\frac{1}{\sqrt{nS}}\>\sum_{j=1}^{n} {\,}|\phi_{j}\rangle\otimes|\phi_{j}^{*}\rangle
\, , \label{phidf}\\
|\Psi_{k}\rangle
&=\frac{1}{\sqrt{n-nc}}\>\sum_{j=1}^{n} {\,}\omega_{n}^{k(j-1)}|\phi_{j}\rangle\otimes|\phi_{j}^{*}\rangle
\, , \label{psidf}
\end{align}
where $k=1,\ldots,n-1$ and $\omega_{n}$ is a primitive $n$-th root of
unity. Then the vectors (\ref{phidf}) and (\ref{psidf}) form an
orthonormal set in the space $\hh_{d}\otimes\hh_{d}$.
\end{aropn1}

{\bf Proof.} First, we aim to show that the vectors
(\ref{phidf})--(\ref{psidf}) are mutually orthogonal. Up to a
factor, the inner product $\langle\Psi_{0}|\Psi_{k}\rangle$ with
$k\neq0$ is represented as
\begin{equation}
\sum_{i=1}^{n} \sum_{j=1}^{n} {\,}\omega_{n}^{k(j-1)} \bigl|\langle\phi_{i}|\phi_{j}\rangle\bigr|^{2}
=\sum_{j=1}^{n} \omega_{n}^{k(j-1)}+c\sum_{\substack{i,j=1 \\ i\neq{j}}}^{n} \omega_{n}^{k(j-1)}
\, . \label{php1}
\end{equation}
By the definition of $\omega_{n}$, one has $\sum_{j=1}^{n}
\omega_{n}^{k(j-1)}=0$ for $k=1,\ldots,n-1$. We assign this zero sum
by the factor $c$, whence the right-hand side of (\ref{php1})
becomes
\begin{equation}
c\,\sum_{i=1}^{n}\sum_{j=1}^{n} {\,}\omega_{n}^{k(j-1)}
=nc\,\sum_{j=1}^{n} {\,}\omega_{n}^{k(j-1)}=0
\, . \nonumber
\end{equation}
Up to a common factor, we express $\langle\Psi_{q}|\Psi_{k}\rangle$
with $k,q\neq0$ as
\begin{equation}
\sum_{i=1}^{n}\sum_{j=1}^{n} {\,}\omega_{n}^{-q(i-1)} \omega_{n}^{k(j-1)}
\bigl|\langle\phi_{i}|\phi_{j}\rangle\bigr|^{2}=
\sum_{j=1}^{n} {\,}\omega_{n}^{(k-q)(j-1)}+
c\sum_{\substack{i,j=1 \\ i\neq{j}}}^{n} \omega_{n}^{k(j-1)-q(i-1)}
\, . \label{psp1}
\end{equation}
For $q\neq{k}$, the first sum in the right-hand side of (\ref{psp1})
is zero. Multiplying it by $c$, one reduces (\ref{psp1}) to the form
\begin{equation}
c\,\sum_{i=1}^{n} {\,}\omega_{n}^{-q(i-1)} \sum_{j=1}^{n} {\,}\omega_{n}^{k(j-1)}=0
\, . \nonumber
\end{equation}
To check the normalization, we note that
\begin{equation}
\sum_{i=1}^{n} \sum_{j=1}^{n} {\,}\bigl|\langle\phi_{i}|\phi_{j}\rangle\bigr|^{2}
=n+(n^{2}-n)c=nS
\, . \label{phn1}
\end{equation}
The inverse square root of (\ref{phn1}) stands right before the sum
in the right-hand side of (\ref{phidf}). Substituting $q=k$ in
(\ref{psp1}) finally leads to
\begin{equation}
n+c\,\sum_{i=1}^{n}\sum_{j=1}^{n} {\,}\omega_{n}^{k(j-i)}-nc
=n-nc
\, , \label{psps12}
\end{equation}
whence the normalization factor of (\ref{psidf}) follows.
$\blacksquare$

The statement of Proposition \ref{pnn1} generalizes one of the
results of \cite{rastmubs}. For a SIC-POVM, we have $n=d^{2}$, so
that the vectors (\ref{phidf}) and (\ref{psidf}) form an orthonormal
basis and induce a unitary matrix of size $d^{2}$. It is instructive
to consider explicitly an example of such basis. The simplest
construction in dimension two contains the kets
\begin{equation}
|\phi_{1}\rangle=|0\rangle
\, , \quad
\sqrt{3}\,|\phi_{2}\rangle=|0\rangle+\sqrt{2}\,|1\rangle
\, , \quad
\sqrt{3}\,|\phi_{3}\rangle=|0\rangle+\sqrt{2}\,\omega_{3}|1\rangle
\, , \quad
\sqrt{3}\,|\phi_{4}\rangle=|0\rangle+\sqrt{2}\,\omega_{3}^{*}|1\rangle
\, , \label{pphis}
\end{equation}
where $\omega_{3}=\exp(\iu2\pi/3)$. Calculating the vectors
(\ref{phidf})--(\ref{psidf}) in the canonical basis leads to the
unitary matrix
\begin{equation}
\frac{1}{2\sqrt{3}}
\begin{pmatrix}
\sqrt{6} & \sqrt{2} & \sqrt{2} & \sqrt{2} \\
0 & (1+\sqrt{3})\,\omega_{6} & -2\,\omega_{6} & (1-\sqrt{3})\,\omega_{6} \\
0 & (1-\sqrt{3})\,\omega_{6}^{*} & -2\,\omega_{6}^{*} & (1+\sqrt{3})\,\omega_{6}^{*} \\
\sqrt{6} & -\sqrt{2} & -\sqrt{2} & -\sqrt{2} \\
\end{pmatrix}
 . \label{matr44}
\end{equation}
It is shown in \cite{abfg19} that associated to a SIC in dimension
$d$ there exist a projector in dimension $d^{2}$ and a Hermitian
Hadamard matrix of size $d^{2}$. A square matrix $\mm$ of size $n$
consisting of unimodular entries is called a Hadamard one, when
$\mm^{\dagger}\mm=n\pen_{n}$ \cite{tz06}. Its rescaling leads to a
unitary matrix all of whose elements have the same absolute value.
The matrix built in corollary 2 of \cite{abfg19} is a Hermitian
Hadamard one. In contrast, the matrix (\ref{matr44}) is not
proportional to a Hermitian one.

To formulate entropic uncertainty relations for an ETF-based
measurement, we proceed to estimating the index of coincidence
(\ref{indef}) from above. Let us extend one of the derivations
presented in \cite{rastmubs} for a SIC-POVM.

\newtheorem{aropn2}[aropn1]{Proposition}
\begin{aropn2}\label{pnn2}
Let $n$ unit kets $|\phi_{j}\rangle$ form an ETF in $\hh_{d}$, and
let POVM $\cle$ be assigned to this frame by (\ref{sres1}). For the
given density matrix $\bro$, it holds that
\begin{equation}
I(\cle;\bro)\leq
\frac{Sc+(1-c)\,\tr(\bro^{2})}{S^{2}}
\ , \label{inbr}
\end{equation}
where $S$ and $c$ obey (\ref{abeq}).
\end{aropn2}

{\bf Proof.} For arbitrary operator $\km$ on $\hh_{d}$ and ket
$|\psi\rangle$, one has
\begin{equation}
\sum_{i=1}^{n}\sum_{j=1}^{n}{\,}\langle\phi_{i}|\km|\phi_{j}\rangle\langle\phi_{j}|\phi_{i}\rangle=\frac{n^{2}}{d^{2}}\>\tr(\km)
 \label{con1}
\end{equation}
and
\begin{equation}
\sum_{j=1}^{n}{\,}|\phi_{j}\rangle\langle\phi_{j}|\psi\rangle=\frac{n}{d}\>|\psi\rangle
\, . \label{con2}
\end{equation}
They follow by substituting the resolution of the identity in
$\tr(\pen_{d}\km\pen_{d})=\tr(\km)$ and
$\pen_{d}|\psi\rangle=|\psi\rangle$, respectively. Let us represent
$\bro\otimes\pen_{d}|\Psi_{0}\rangle$ in the form
\begin{equation}
\bro\otimes\pen_{d}|\Psi_{0}\rangle=\sum_{k=0}^{n-1}{\,}a_{k}|\Psi_{k}\rangle+|\Theta\rangle
\, , \label{hteta}
\end{equation}
where $\langle\Theta|\Psi_{k}\rangle=0$ for all $k=0,1,\ldots,n-1$.
It follows from (\ref{phidf}) and (\ref{con1}) that
\begin{equation}
a_{0}=\langle\Psi_{0}|\bro\otimes\pen_{d}|\Psi_{0}\rangle
=\frac{1}{nS}{\,}\sum_{i=1}^{n}\sum_{j=1}^{n} {\,}\langle\phi_{i}|\bro|\phi_{j}\rangle\langle\phi_{j}|\phi_{i}\rangle
=\frac{n}{Sd^{2}}=\frac{1}{d}
\ . \label{bpph}
\end{equation}
Using (\ref{con2}), for $k\neq0$, we write the coefficient
\begin{align}
a_{k}=\langle\Psi_{k}|\bro\otimes\pen_{d}|\Psi_{0}\rangle
&=\frac{1}{n\sqrt{S-Sc}}\>\sum_{i=1}^{n}\sum_{j=1}^{n} {\,}
\omega_{n}^{-k(i-1)}\langle\phi_{i}|\bro|\phi_{j}\rangle\langle\phi_{j}|\phi_{i}\rangle
\nonumber\\
&=\frac{S^{2}}{n\sqrt{S-Sc}}\>\sum_{i=1}^{n}{\,}\omega_{n}^{-k(i-1)}p_{i}
\, . \label{akdf}
\end{align}
Similarly to (\ref{bpph}), one also calculates
\begin{equation}
\langle\Psi_{0}|(\bro\otimes\pen_{d})^{2}|\Psi_{0}\rangle
=\frac{1}{nS}{\,}\sum_{i=1}^{n}\sum_{j=1}^{n} {\,}\langle\phi_{i}|\bro^{2}|\phi_{j}\rangle\langle\phi_{j}|\phi_{i}\rangle
=\frac{\tr(\bro^{2})}{d}\geq\frac{1}{d^{2}}+\sum_{k=1}^{n-1}{\,}a_{k}^{*}a_{k}
\,. \label{bpph2}
\end{equation}
Due to (\ref{akdf}) and
$\sum_{k=1}^{n-1}\omega_{n}^{k(i-j)}=n\delta_{ij}-1$, we obtain
\begin{equation}
\sum_{k=1}^{n-1}{\,}a_{k}^{*}a_{k}=\frac{S^{3}}{n^{2}(1-c)}\>\sum_{i=1}^{n}\sum_{j=1}^{n}{\,}p_{i}p_{j}\sum_{k=1}^{n-1}{\,}\omega_{n}^{k(i-j)}
=\frac{n^{2}I(\cle;\bro)-n}{d^{3}(1-c)}
\ . \label{sumaka}
\end{equation}
Combining (\ref{bpph2}) with (\ref{sumaka}) leads to
\begin{equation}
n^{2}I(\cle;\bro)\leq{n}-d(1-c)+d^{2}(1-c)\,\tr(\bro^{2})
=ncd+d^{2}(1-c)\,\tr(\bro^{2})
\, , \label{n2in}
\end{equation}
since $n-d(1-c)=ncd$ in line with (\ref{abeq}). Dividing
(\ref{n2in}) by $n^{2}$ gives (\ref{inbr}) due to $S=n/d$.
$\blacksquare$

For the maximally mixed state $\bro_{*}=\pen_{d}/d$, the inequality
(\ref{inbr}) is saturated. By $1-c=S-nc$ and $Sd=n$, the
right-hand side of (\ref{inbr}) reads as
\begin{equation}
\frac{Scd+1-c}{S^{2}d}=\frac{nc+S-nc}{S^{2}d}=\frac{1}{n}
\ . \label{inbrcm}
\end{equation}
It is equal to $I(\cle;\bro_{*})=1/n$ as follows from
$p_{j}(\cle;\bro_{*})=1/n$ for all $j$. Also, the inequality
(\ref{inbr}) is saturated with any pure state $|\phi_{j}\rangle$
taken from the frame kets. Combining this with (\ref{bpph2}) leads
us to a conclusion. Namely, the inequality (\ref{inbr}) is saturated
for each density matrix of the form
\begin{equation}
\bsg=\sum_{j=1}^{n}\mu_{j}\,|\phi_{j}\rangle\langle\phi_{j}|
\, , \label{nuj}
\end{equation}
where non-negative weights $\mu_{j}$ sum to $1$. Indeed, if two
density matrices on $\hh_{d}$ are such that $|\Theta\rangle$ in
(\ref{hteta}) is zero, then their convex combination also has this
property. Then the sign of inequality in (\ref{bpph2}) is replaced
with equality. It is not difficult to check the made conclusion
immediately. Let $\vark=\sum_{j=1}^{n}\mu_{j}^{2}$, then
calculations show
\begin{align}
S{\;\!}p_{j}(\cle;\bsg)&=\mu_{j}+c(1-\mu_{j})
\, , \label{pir}\\
S^{2}I(\cle;\bsg)&=(1-c)^{2}\vark+2c(1-c)+nc^{2}
 \label{rbr1}
\end{align}
and
\begin{equation}
\tr(\bsg^{2})=(1-c)\vark+c
\, . \label{rbr2}
\end{equation}
Substituting (\ref{rbr1}) and (\ref{rbr2}) in (\ref{inbr}) with
the equality sign and eliminating $\vark$ gives the formula
equivalent to (\ref{abeq}).

Let us discuss briefly the case of SIC-POVMs. For $n=d^{2}$, $S=d$
and $c=(d+1)^{-1}$, the result (\ref{inbr}) is rewritten as
\begin{equation}
I(\cle;\bro)=\frac{1+\tr(\bro^{2})}{d(d+1)}
 \label{inbrsic}
\end{equation}
with the probabilities
$p_{j}(\cle;\bro)=d^{-1}\langle\phi_{j}|\bro|\phi_{j}\rangle$. The
equality holds since the set of kets (\ref{phidf}) and (\ref{psidf})
is complete here. As was mentioned, the index of coincidence
(\ref{inbrsic}) was calculated in \cite{rastmubs}. The inequality
(\ref{inbr}) reflects a non-trivial inner structure of ETFs. Hence,
various uncertainty relations for the corresponding POVMs
immediately follow.

\section{Fine-grained uncertainty relations}\label{sec4}

For a pair of observables, uncertainty relations of the
Landau--Pollak type are formulated in terms of the two maximal
probabilities. Its quantum-mechanical interpretation was mentioned
in \cite{maass88}, since the original formulation of Landau \&
Pollak \cite{landau61} was focused on signal analysis. Relations of
the Landau--Pollak type can also be treated as an example of
fine-grained uncertainty relations. The authors of \cite{oppew10}
emphasized the role of such relations dealing with a particular
combination of the outcomes. When the number of measurement outcomes
exceeds dimensionality, a non-trivial upper bound holds already for
a single probability \cite{rastmubs}. For the given index of
coincidence, the maximal probability can be estimated from above as
described in \cite{rastmubs}. Using (\ref{inbr}), we then have
\begin{align}
\underset{j}{\max}\>p_{j}(\cle;\bro)
&\leq\frac{1}{n}
\left(
1+\sqrt{n-1}\,\sqrt{nI(\cle;\bro)-1}
\,\right)
\label{maxjp}\\
&\leq\frac{1}{nS}
\left(
S+\sqrt{n-1}\,\sqrt{nSc+n(1-c)\,\tr(\bro^{2})-S^{2}}
\,\right)
\nonumber\\
&=\frac{1}{nS}
\left(
S+\sqrt{(n-1)(1-c)}\,\sqrt{n\,\tr(\bro^{2})-S}
\,\right)
 . \label{maxjp1}
\end{align}
This inequality is saturated for the maximally mixed state
$\bro_{*}=\pen_{d}/d$. Indeed, we clearly have
$n\,\tr(\bro_{*}^{2})=S$. It is not so obvious that it is
saturated for any of the frame states, when the right-hand side
of (\ref{maxjp1}) reduces to $d/n$. Of course, the
state-independent inequality
\begin{equation}
\underset{j}{\max}\>p_{j}(\cle;\bro)\leq\frac{d}{n}
 \label{maxjdn}
\end{equation}
directly follows from (\ref{sres1}). In the simplest case of
orthonormal bases, when $n=d$ and $c=0$, the inequality
(\ref{maxjp}) is saturated for all basis states. In general, the
inequality (\ref{maxjp1}) is not tight for density matrices of the
form (\ref{nuj}), though they saturate (\ref{inbr}). In fact, the
inequality (\ref{maxjp}) becomes equality only if all the
probabilities except possibly the maximal one are equal. Let $p_{1}$
denote the maximal probability. By convexity of the
function $\xi\mapsto\xi^{2}$, we have
\begin{equation}
I(p_{1},\ldots,p_{n})\geq{p}_{1}^{2}+\frac{(1-p_{1})^{2}}{n-1}=\frac{np_{1}^{2}-2p_{1}+1}{n-1}
\ , \label{ip1n}
\end{equation}
with equality if and only if $p_{j}=(n-1)^{-1}(1-p_{1})$ for all
$j=2,\ldots,n$. The latter implies
\begin{equation}
(n-1)\bigl[nI(p_{1},\ldots,p_{n})-1\bigr]=n^{2}p_{1}^{2}-2np_{1}+n-(n-1)
=(np_{1}-1)^{2}
 \label{sqnp}
\end{equation}
and equality in (\ref{maxjp}) due to $np_{1}\geq1$. In view of
(\ref{pir}), the inequalities (\ref{maxjp}) and (\ref{maxjp1}) are
both saturated for (\ref{nuj}) provided that all the weights except
possibly the maximal one are equal.

It is instructive to discuss a contrast case with one half of
weights $\mu_{j}$ equal to $2n^{-1}$ and zero others (we assume $n$
to be even that is not principal for sufficiently large values).
Restricting a consideration to $n=d^{2}$, we obtain
\begin{equation}
p_{j}(\cle;\bsg)=\frac{d+2}{d^{2}(d+1)}
 \label{ppj}
\end{equation}
for $j$ with $\mu_{j}=2d^{-2}$ and $p_{j}(\cle;\bsg)=\bigl(d(d+1)\bigr)^{-1}$
for $j$ with $\mu_{j}=0$. In this case, the left-hand side of (\ref{maxjp})
is represented by (\ref{ppj}). It also holds that
\begin{equation}
\sqrt{nI(\cle;\bsg)-1}=\frac{1}{d+1}
 \nonumber
\end{equation}
and
\begin{equation}
\frac{1}{n}
\left(
1+\sqrt{n-1}\,\sqrt{nI(\cle;\bsg)-1}
\,\right)=
\frac{1}{d^{2}}\,\biggl(1+\sqrt{\frac{d-1}{d+1}}\,\biggr)
\, . \label{ubon}
\end{equation}
Comparing (\ref{ppj}) with (\ref{ubon}) in our example shows the
following. For sufficiently large $d$, the right-hand side of
(\ref{maxjp}) behaves like the doubled maximal probability. On the
other hand, this estimation from above approximately equal to
$2d^{-2}$ is essentially better than obvious estimate $d^{-1}$ based
on (\ref{maxjdn}).

The structure of an ETF allows us to derive a family of fine-grained
uncertainty relations, though in state-independent formulation. Let
$\cli=\{i_{1},\ldots,i_{m}\}$ be a non-empty subset of the set
$\{1,\ldots,n\}$, and let
\begin{equation}
\mtgm_{\cli}=
\bigl(|\phi_{i_{1}}\rangle\,\cdots\,|\phi_{i_{m}}\rangle\bigr)
\, . \label{gncd}
\end{equation}
This is a $d\times{m}$ matrix written in terms of $m$ columns.
Recall also that the spectral norm $\|\km\|_{\infty}$ is defined as
the square root of the maximum eigenvalue of $\km^{\dagger}\km$. The
following statement takes place.

\newtheorem{aropn23}[aropn1]{Proposition}
\begin{aropn23}\label{pnn23}
Let $n$ unit kets $|\phi_{j}\rangle$ form an ETF in $\hh_{d}$, and let
POVM $\cle$ be assigned to this frame by (\ref{sres1}). For the
given density matrix $\bro$ and particular combination
$\cli=\{i_{1},\ldots,i_{m}\}$ of outcomes, it holds that
\begin{equation}
\sum_{i\in\cli}p_{i}(\cle;\bro)\leq
\frac{d}{n}\,\bigl\|\mtgm_{\cli}^{\dagger}\mtgm_{\cli}\bigr\|_{\infty}
\ , \label{fgunr}
\end{equation}
where $\mtgm_{\cli}$ is defined by (\ref{gncd}).
\end{aropn23}

{\bf Proof.} It is sufficient to prove (\ref{fgunr}) for pure
states. For any unit ket $|\psi\rangle$, the sum of probabilities
reads as
\begin{equation}
\frac{d}{n}\,\sum_{i\in\cli}\langle\psi|\phi_{i}\rangle\langle\phi_{i}|\psi\rangle=
\frac{d}{n}\,\langle\psi|\,\mtgm_{\cli}\mtgm_{\cli}^{\dagger}|\psi\rangle
\leq\frac{d}{n}\,\bigl\|\mtgm_{\cli}\mtgm_{\cli}^{\dagger}\bigr\|_{\infty}
=\frac{d}{n}\,\bigl\|\mtgm_{\cli}^{\dagger}\mtgm_{\cli}\bigr\|_{\infty}
\ . \label{fgunr1}
\end{equation}
The final step uses the fact that $\mtgm_{\cli}\mtgm_{\cli}^{\dagger}$ and
$\mtgm_{\cli}^{\dagger}\mtgm_{\cli}$ have the same non-zero eigenvalues.
$\blacksquare$

The above reasons are like the approach proposed in \cite{rastsep}
to extend uncertainty relations of the papers \cite{mupd,musm} to
POVMs with rank-one elements. The statement of Proposition
\ref{pnn23} is a kind of fine-grained uncertainty relations for an
ETF-based measurement. Of course, similar reasons hold for arbitrary
POVM with rank-one elements. On the other hand, the result can
hardly be useful without explicit knowledge of the matrix elements.
Nonetheless, the structure of an ETF is such that $m\times{m}$
matrix $\mtgm_{\cli}^{\dagger}\mtgm_{\cli}$ has ones on the main
diagonal and off-diagonal entries of the form
$\langle\phi_{i}|\phi_{j}\rangle=\sqrt{c}\,\exp(-\iu\theta_{ij})$.
For $m=2$, we get
\begin{equation}
\underset{i\neq{j}}{\max}\,\bigl\{p_{i}(\cle;\bro)+p_{j}(\cle;\bro)\bigr\}=\frac{d+\sqrt{c}\,d}{n}
\ . \label{mxijdn}
\end{equation}
Indeed, the spectral norm of a positive matrix is equal to the
maximum of its eigenvalues. The right-hand side of (\ref{mxijdn}) is
product of $d/n$ and the maximum of eigenvalues $1\pm\sqrt{c}$
calculated for
\begin{equation}
\begin{pmatrix}
1 & \sqrt{c}\,e^{-\iu\theta}\, \\
\sqrt{c}\,e^{\iu\theta} & 1
\end{pmatrix}
 . \label{ggdr}
\end{equation}
Here, the phase factor corresponds to
$\langle\phi_{j}|\phi_{i}\rangle=\sqrt{c}\,e^{\iu\theta}$. Thus, we
obtain a state-independent uncertainty relation with the sum of
two probabilities. It is easy to check that the right-hand side of
(\ref{mxijdn}) is reached with the ket
\begin{equation}
\frac{|\phi_{i}\rangle+e^{\iu\theta}|\phi_{j}\rangle}{\sqrt{2+2\sqrt{c}}}
\ . \nonumber
\end{equation}
Thus, the state-independent formulation (\ref{mxijdn}) cannot be
improved under chosen circumstances.

The consideration is more complicated for $m=3$. By
$|\phi_{i}\rangle$, $|\phi_{j}\rangle$ and $|\phi_{k}\rangle$, we
denote three columns of the matrix (\ref{gncd}). The eigenvalues of
$\mtgm_{\cli}^{\dagger}\mtgm_{\cli}$ obey characteristic equation
\begin{equation}
\lambda^{3}-3\lambda^{2}+(3-3c)\lambda-(1-3c+2c^{3/2}\cos\gamma)=0
\, , \qquad
\gamma=\theta_{ij}+\theta_{jk}+\theta_{ki}
\, . \label{charq}
\end{equation}
By the standard approach, we rewrite (\ref{charq}) as
$(\lambda-1)^{3}-3c(\lambda-1)-2c^{3/2}\cos\gamma=0$ and substitute
$u+v=\lambda-1$, $uv=c$. It then follows that both the terms $u^{3}$
and $v^{3}$ satisfy quadratic equation
\begin{equation}
\xi^{2}-2\xi{c}^{3/2}\cos\gamma+c^{3}=0
 \nonumber
\end{equation}
with the roots $\xi_{\pm}=c^{3/2}\exp(\pm\iu\gamma)$. Doing usual
algebra finally gives the set of eigenvalues
\begin{equation}
\left\{1+2\sqrt{c}\,\cos\frac{\gamma}{3}\>,1+2\sqrt{c}\,\cos\Bigl(\frac{\gamma}{3}\pm\frac{2\pi}{3}\Bigr)\right\}
 . \nonumber
\end{equation}
None of these eigenvalues exceeds $1+2\sqrt{c}$, whence
\begin{equation}
\underset{i\neq{j}\neq{k}\neq{i}}{\max}\bigl\{p_{i}(\cle;\bro)+p_{j}(\cle;\bro)+p_{k}(\cle;\bro)\bigr\}\leq\frac{d+2\sqrt{c}\,d}{n}
\ . \label{m3dn}
\end{equation}
In general, this inequality is not tight, since we replaced exact
values of cosines with $1$. On the other hand, the exact phases are
known in each of concrete cases. Using these phases allows one to
improve (\ref{m3dn}). We refrain from exemplifying this possibility
here. Being state-independent, the inequalities (\ref{mxijdn}) and
(\ref{m3dn}) can be applied to detect non-classical correlations.

\section{Entropic uncertainty relations}\label{sec5}

Entropic uncertainty relations are a prosperous alternative to more
traditional approach \cite{deutsch,kraus87}. For a general
discussion of entropic uncertainty relations and their role, see the
review \cite{cbtw17} and references therein. To express uncertainty
relations, we will use the R\'{e}nyi and Tsallis entropies. For
$\alpha>0\neq1$, the R\'{e}nyi $\alpha$-entropy is expressed as
\cite{renyi61}
\begin{equation}
R_{\alpha}(p)=\frac{1}{1-\alpha}\>\ln\!\left(\sum\nolimits_{j} p_{j}^{\alpha}
\right)
 . \label{renent}
\end{equation}
This quantity reduces to the Shannon
entropy $H_{1}(p)=-\sum_{j}p_{j}\ln{p}_{j}$ in the limit $\alpha\to1$. For $\alpha=0$
we have logarithm of the number of non-zero probabilities, whereas
the limit $\alpha\to\infty$ gives the min-entropy
\begin{equation}
R_{\infty}(p)=-\ln\bigl(\max p_{j}\bigr)
\, . \label{mnen}
\end{equation}
For $\alpha>0\neq1$, the Tsallis $\alpha$-entropy is defined by
\cite{tsallis}
\begin{equation}
H_{\alpha}(p)=\frac{1}{1-\alpha}\left(\sum\nolimits_{j} p_{j}^{\alpha}
- 1 \right)
=-\sum\nolimits_{j}p_{j}^{\alpha}\ln_{\alpha}(p_{j})
\, . \label{tsaent}
\end{equation}
Here, the $\alpha$-logarithm of positive $\xi$ is expressed as
\begin{equation}
\ln_{\alpha}(\xi)=
\begin{cases}
 \frac{\xi^{1-\alpha}-1}{1-\alpha}\>, & \text{ for } 0<\alpha\neq1 \, , \\
 \ln\xi\, , & \text{ for } \alpha=1 \,.
\end{cases}
\nonumber
\end{equation}
By $R_{\alpha}(\cle;\bro)$ and $H_{\alpha}(\cle;\bro)$, we
respectively mean the entropies (\ref{renent}) and (\ref{tsaent})
computed with the probabilities (\ref{prbj}). By construction, the
Tsallis is concave with respect to probability distributions. The
R\'{e}nyi $\alpha$-entropy is certainly concave for $0<\alpha\leq1$
\cite{ja04}. For $\alpha>1$, the answer depends on dimensionality of
probability distributions \cite{bz17}.

Entropic uncertainty relations for a POVM with elements
(\ref{sres1}) follow from the inequality (\ref{inbr}). Let us begin
with uncertainty relations in terms of R\'{e}nyi entropies. Combining
(\ref{maxjp1}) with (\ref{mnen}) leads to the min-entropy uncertainty
relation,
\begin{equation}
R_{\infty}(\cle;\bro)\geq\ln(nS)
-\ln\!\left(
S+\sqrt{(n-1)(1-c)}\,\sqrt{n\,\tr(\bro^{2})-S}
\,\right)
 , \label{unrr1}
\end{equation}
where $S$ and $c$ are defined by (\ref{abeq}). In the case
$n=d^{2}$ with a SIC-POVM, the inequality
(\ref{unrr1}) reduces to one of the results of \cite{rastmubs}.
Substituting $\alpha=2$ in (\ref{renent}) gives the collision
entropy $R_{2}(\cle;\bro)=-\ln{I}(\cle;\bro)$. Combining the latter
with (\ref{inbr}), we obtain
\begin{align}
R_{2}(\cle;\bro)&\geq2\ln{S}
-\ln\!\left(
Sc+(1-c)\,\tr(\bro^{2})
\right)
\label{unrr2}\\
&\geq\ln\!\left(\frac{n^{2}-n}{d^{2}-2d+n}\right)
 . \label{unrr2p}
\end{align}
Similarly to (\ref{inbr}), the inequality (\ref{unrr2}) is saturated
for states of the form (\ref{nuj}). If all the weights except
possibly the maximal one are equal then such states also saturate
(\ref{unrr1}). As was shown in \cite{rastosid}, the R\'{e}nyi
$\alpha$-entropy of order $\alpha\in[2,\infty]$ satisfies
\begin{equation}
R_{\alpha}(p)\geq
\frac{\alpha-2}{\alpha-1}\,R_{\infty}(p)+\frac{1}{\alpha-1}\,R_{2}(p)
\, . \label{al2eq}
\end{equation}
Combining the above inequalities, we have arrived at a conclusion.

\newtheorem{aropn3}[aropn1]{Proposition}
\begin{aropn3}\label{pnn3}
Let $n$ unit kets $|\phi_{j}\rangle$ form an ETF in $\hh_{d}$, and
let POVM $\cle$ be assigned to this frame by (\ref{sres1}). For the
given density matrix $\bro$ and $\alpha\in[2,\infty]$, it holds that
\begin{align}
R_{\alpha}(\cle;\bro)&\geq
\frac{\alpha\ln{S}+(\alpha-2)\ln{n}-\ln\!\left(
Sc+(1-c)\,\tr(\bro^{2})
\right)}{\alpha-1}
\nonumber\\
&-\frac{\alpha-2}{\alpha-1}\,
\ln\!\left(
S+\sqrt{(n-1)(1-c)}\,\sqrt{n\,\tr(\bro^{2})-S}
\,\right)
 \label{unrr12}\\
&\geq\ln{n}
-\frac{(\alpha-2)\ln{d}}{\alpha-1}+
\frac{1}{\alpha-1}\,
\ln\!\left(
\frac{n-1}{d^{2}-2d+n}
\right)
 . \label{unrr12p}
\end{align}
\end{aropn3}

The statement of Proposition \ref{pnn3} gives R\'{e}nyi-entropy
uncertainty relations for the POVM assigned to an ETF. These relations
reflect features of ETFs with the use of (\ref{inbr}) in
combination with some properties of R\'{e}nyi entropies. It is well
known that the entropy (\ref{renent}) cannot increase with growth of
$\alpha$. Hence, the inequality (\ref{unrr2}) remains valid with the
R\'{e}nyi $\alpha$-entropy of order $\alpha\in[0,2]$. It would be
interesting to obtain for this interval more precise inequalities.
Apparently, other ways of derivation should be considered here.

To derive uncertainty relations in terms of Tsallis entropies, we
apply the corresponding methods of \cite{rastmubs}. The following
statement takes place.

\newtheorem{aropn4}[aropn1]{Proposition}
\begin{aropn4}\label{pnn4}
Let $n$ unit kets $|\phi_{j}\rangle$ form an ETF in $\hh_{d}$, and
let POVM $\cle$ be assigned to this frame by (\ref{sres1}). For the
given density matrix $\bro$ and $\alpha\in(0,2]$, it holds that
\begin{equation}
H_{\alpha}(\cle;\bro)\geq
\ln_{\alpha}\!\left(
\frac{S^{2}}{Sc+(1-c)\,\tr(\bro^{2})}
\right)
 . \label{unhh02}
\end{equation}
The state-independent inequality is posed as
\begin{equation}
H_{\alpha}(\cle;\bro)\geq
\ln_{\alpha}\!\left(
\frac{n^{2}-n}{d^{2}-2d+n}
\right)
 . \label{unhh02p}
\end{equation}
\end{aropn4}

{\bf Proof.}
For $0<\alpha\leq2$, one writes
\begin{equation}
H_{\alpha}(p)\geq\ln_{\alpha}\!\left(\frac{1}{I(p)}\right)
 . \label{conc1}
\end{equation}
The latter holds by applying Jensen's inequality to the function
$\xi\mapsto\ln_{\alpha}(1/\xi)$ \cite{rastmubs}. Since this function
decreases, combining (\ref{inbr}) with (\ref{conc1}) implies
(\ref{unhh02}). The result (\ref{unhh02p}) immediately follows due
to $\tr(\bro^{2})\leq1$.
$\blacksquare$

The statement of Proposition \ref{pnn4} gives Tsallis-entropy
uncertainty relations for the measurement assigned to an ETF. There
are especially interesting values of $\alpha$. For $\alpha=2$, the
Tsallis entropy is equal to $1$ minus the index of coincidence.
Here, the inequality (\ref{unhh02}) is saturated for all states of
the form (\ref{nuj}). Moreover, for a SIC-POVM we always have
\begin{equation}
H_{2}(\cle;\bro)=
1-\frac{1+\tr(\bro^{2})}{d(d+1)}
\ . \label{unhh02s}
\end{equation}
In the case $\alpha=1$, the inequality (\ref{unhh02}) can be
improved due to the results of the paper \cite{harr2001}. Its
authors considered information diagrams that represent the Shannon
entropy as a function of the index of coincidence. It is seen from
(\ref{conc1}) that $H_{1}(p)\geq-\ln{I}(p)$. Here, the smooth line
$\xi\mapsto-\ln\xi$ can be replaced with the polygonal line
connecting the points $(1/q,\ln{q})$ with integer $q=1,\ldots,n$.
Each of them corresponds to a uniform distribution with the
corresponding number of non-zero probabilities. Thus, the final
inequality reads as
\begin{equation}
H_{1}(\cle;\bro)\geq\underset{1\leq{q}\leq{n}-1}{\max}
\left\{\ln(q+1)+q\ln\bigl(1+q^{-1}\bigr)-q(q+1)\ln\bigl(1+q^{-1}\bigr)
\,\frac{Sc+(1-c)\,\tr(\bro^{2})}{S^{2}}
\right\}
 . \label{unsh02}
\end{equation}
Extending the methods of the paper \cite{harr2001} to generalized
entropies is an open question. Entropic uncertainty relations
are a tool for deriving several kinds of criteria used in quantum
information processing \cite{cbtw17}. Such criteria are of interest,
since ETFs are easier to construct than SIC-POVMs. In more detail,
applications of the derived relations will be discussed in the next
section.

Another reason to use Tsallis-entropy uncertainty relations deals
with the case of detection inefficiencies. To the given
efficiency $\eta\in[0;1]$ and probability distribution $\{p_{j}\}$,
we assign a distorted distribution
\begin{equation}
p_{j}^{(\eta)}=\eta{\,}p_{j}
\, , \qquad
p_{\varnothing}^{(\eta)}=1-\eta
\, . \label{pdds}
\end{equation}
The probability $p_{\varnothing}^{(\eta)}$ corresponds to the
no-click event. It is easy to check that \cite{rastqqt}
\begin{equation}
H_{\alpha}{\bigl(p^{(\eta)}\bigr)}=\eta^{\alpha}H_{\alpha}(p)+h_{\alpha}(\eta)
\, , \label{qtlm0}
\end{equation}
where the binary entropy
$h_{\alpha}(\eta)=(1-\alpha)^{-1}\bigl[\eta^{\alpha}+(1-\eta)^{\alpha}-1\bigr]$.
For $\alpha\in(0,2]$, the Tsallis $\alpha$-entropy calculated with
(\ref{pdds}) satisfies
\begin{equation}
H_{\alpha}^{(\eta)}(\cle;\bro)\geq
\eta^{\alpha}\ln_{\alpha}\!\left(
\frac{S^{2}}{Sc+(1-c)\,\tr(\bro^{2})}
\right)
+h_{\alpha}(\eta)
\, . \label{punhh02}
\end{equation}
The latter immediately follows from (\ref{unhh02}) and
(\ref{qtlm0}). For the Shannon entropy, we also have
\begin{align}
H_{1}^{(\eta)}(\cle;\bro)&\geq
\eta\underset{1\leq{q}\leq{n}-1}{\max}
\left\{\ln(q+1)+q\ln\bigl(1+q^{-1}\bigr)-q(q+1)\ln\bigl(1+q^{-1}\bigr)
\,\frac{Sc+(1-c)\,\tr(\bro^{2})}{S^{2}}
\right\}
\nonumber\\
&+h_{1}(\eta)
\, . \label{punsh02}
\end{align}
In other words, entropies of the actual probability distribution
reflect not only quantum uncertainties, but also an additional
uncertainty inserted by detectors. The right-hand sides of
(\ref{punhh02}) and (\ref{punsh02}) also allow us to estimate a
required amount of detector efficiency. In both the formulae, the
first term should be sufficiently large in comparison with the
corresponding binary entropy.

A utility of relations with parametric dependence has been
exemplified in \cite{maass88}. Varying entropic parameters is
important in asking for possible values of unknown characteristics
at the given ones. Let us discuss this for relations of
the form (\ref{unhh02}) in application to a SIC-POVM. Then the index
of coincidence is connected with the state purity by
(\ref{inbrsic}). If $n-2$ probabilities are given, then the
remaining two are determined from the normalization condition and
the value of $H_{2}(\cle;\bro)$ calculated in line with
(\ref{unhh02s}). When $\alpha$ deviates from $2$, ambiguity in
determining unknown probabilities increases. A special role of the
value $\alpha=2$ was already emphasized in \cite{cug18}. In the case
of detection inefficiencies, the picture becomes more complicated,
but a possibility to vary entropic parameters is still important due
to relations of the form (\ref{punhh02}).

\section{Some applications}\label{sec6}

MUBs give an important tool to detect
entanglement \cite{shbah12}. The use of SIC-POVMs in entanglement
detection was briefly discussed in \cite{rastmubs}. This task can
also be approached with measurements assigned to ETFs. By $A$ and
$B$, we denote subsystems of a bipartite system with Hilbert space
$\hh_{AB}=\hh_{A}\otimes\hh_{B}$.
 A bipartite mixed state is called
separable, when its density matrix can be represented as a convex
combination of product states \cite{werner89,zhsl98}. To use ETFs
for entanglement detection, we extend the method considered in
\cite{rastmubs}. Let us consider a bipartite system of two
$d$-dimensional subsystems. To the given ETF, we assign the POVM
$\cln_{AB}$ with elements
\begin{equation}
\frac{d^{2}}{n^{2}}\>
|\phi_{j}\rangle\langle\phi_{j}|\otimes|\phi_{k}^{*}\rangle\langle\phi_{k}^{*}|
\, . \nonumber
\end{equation}
For a density matrix $\bro_{AB}$ of the bipartite system, the
$(j,k)$-probability reads as
\begin{equation}
\frac{d^{2}}{n^{2}}\,\langle\phi_{j}\phi_{k}^{*}|\bro_{AB}|\phi_{j}\phi_{k}^{*}\rangle
\, . \label{phjk1}
\end{equation}
Summing these probabilities over all $j=k$, we obtain the
correlation measure
\begin{equation}
G\bigl(\cln_{AB};\bro_{AB}\bigr)=\frac{d^{2}}{n^{2}}\,\sum_{j=1}^{n}
\langle\phi_{j}\phi_{j}^{*}|\bro_{AB}|\phi_{j}\phi_{j}^{*}\rangle
\, . \label{cmjj}
\end{equation}
For the case of SIC-POVMs, the quantity (\ref{cmjj}) proposed in
\cite{rastmubs}. It is also similar to the mutual predictability
used in \cite{shbah12}. For each product state
$\bro_{A}\otimes\bro_{B}$, the probability (\ref{phjk1}) is a
product of two local probabilities. Using (\ref{inbr}) and the
Cauchy--Schwarz inequality, we then obtain
\begin{equation}
G\bigl(\cln_{AB};\bro_{A}\otimes\bro_{B}\bigr)\leq
\frac{1}{S^{2}}\>\sqrt{Sc+(1-c)\,\tr\bigl(\bro_{A}^{2}\bigr)}
\>\sqrt{Sc+(1-c)\,\tr\bigl(\bro_{B}^{2}\bigr)}
\leq\frac{d^{2}-2d+n}{n^{2}-n}
\ . \label{gab2}
\end{equation}
This inequality takes place for all separable states. Its violation
implies that the tested state is entangled. Using some orthonormal
basis $\bigl\{|b_{i}\rangle\bigr\}$ in $\hh_{d}$, we write a maximally
entangled state
\begin{equation}
|\Phi_{+}\rangle=\frac{1}{\sqrt{d}}\,\sum_{i=1}^{d} |b_{i}\rangle\otimes|b_{i}\rangle
\, . \nonumber
\end{equation}
It is immediate to get
$\langle\phi_{j}\phi_{j}^{*}|\Phi_{+}\rangle=1/\sqrt{d}$, whence the
$(j,j)$-probability is equal to $d/n^{2}$ and
\begin{equation}
G\bigl(\cln_{AB};|\Phi_{+}\rangle\langle\Phi_{+}|\bigr)
=\frac{d}{n}
\ . \label{smen}
\end{equation}
To estimate a region of detectability, we divide the right-hand side
of (\ref{gab2}) by (\ref{smen}). The ratio is calculated as
\begin{equation}
\frac{d-2+n/d}{n-1}<1
\, , \nonumber
\end{equation}
whenever $n>d$. In this way, the most efficient scheme takes place
for $n=d^{2}$, when the used ETF is maximal with reaching a
SIC-POVM. Isotropic states are typically used to test entanglement
criteria. For $\nu\in[0,1]$, one considers the density matrix
\begin{equation}
\vbro_{\;\!\mathrm{iso}}=\nu\,|\Phi_{+}\rangle\langle\Phi_{+}|+\frac{1-\nu}{d^{2}}\,\pen_{d}\otimes\pen_{d}
\, . \label{viso}
\end{equation}
These states are known to be entangled if and only if
$\nu>(d+1)^{-1}$ \cite{hor99}. It can be checked that the
$(j,j)$-probability is equal to $n^{-2}(\nu{d}+1-\nu)$, whence
\begin{equation}
G\bigl(\cln_{AB};\vbro_{\,\mathrm{iso}}\bigr)
=\frac{\nu{d}+1-\nu}{n}
\ . \nonumber
\end{equation}
Combining this with (\ref{gab2}) shows that the violation occurs for
\begin{equation}
\nu>\frac{d-1}{n-1}
\ . \label{smis}
\end{equation}
With a SIC-POVM, the latter gives the maximally wide interval, i.e.,
$\nu>(d+1)^{-1}$. At the same time, ETFs with sufficiently large
number of elements are also of interest to detect entanglement, at
least for isotropic states. No possibility for entanglement
detection we see only for a single orthonormal basis, when $n=d$. In
addition, the use of several ETFs provides a collection of
separability conditions. There is another way to use entropic
uncertainty relations for POVMs in entanglement detection
\cite{rastsep}. In general, this question deserves further study.

For applications in quantum information, uncertainty relations in
the presence of quantum memory are of interest \cite{bccrr10}. Let
us recall the tripartite formulation proposed in \cite{colep14}. It
is equivalent to the bipartite formulation and extends uncertainty
relations of this kind for POVMs. The approach of \cite{colep14}
uses the isometry
$\um_{\cle}:\>\hh_{A}\rightarrow\hh_{X}\otimes\hh_{Y}\otimes\hh_{A}$
defined by
\begin{equation}
\um_{\cle}=\sqrt{\frac{d}{n}}
\>\sum_{j=1}^{n}|x_{j}\rangle\otimes|y_{j}\rangle\otimes|\phi_{j}\rangle\langle\phi_{j}|
\, . \label{}
\end{equation}
By $|x_{j}\rangle$ and $|y_{j}\rangle$, one means kets of the
orthonormal bases in the $n$-dimensional spaces $\hh_{X}$ and
$\hh_{Y}$, respectively. To each tripartite state $\bro_{ABC}$, we
further assign the conditional von Neumann entropies
\begin{align}
\rmh_{1}(X|B)&=\rmh_{1}(\bro_{XB})-\rmh_{1}(\bro_{B})
\, , \label{hxb}\\
\rmh_{1}(X|C)&=\rmh_{1}(\bro_{XC})-\rmh_{1}(\bro_{C})
\, . \label{hxc}
\end{align}
Here, the von Neumann entropy of each $\bro$ is defined as
$\rmh_{1}(\bro)=-\,\tr(\bro\ln\bro)$, and the reduced states are
obtained as the corresponding partial traces of the matrix
\begin{equation}
\bigl(\um_{\cle}\otimes\pen_{BC}\bigr)\bro_{ABC}\bigl(\um_{\cle}^{\dagger}\otimes\pen_{BC}\bigr)
\, . \nonumber
\end{equation}
In the considered case, theorem 5 of \cite{colep14} reads as
\begin{equation}
\rmh_{1}(X|B)+\rmh_{1}(X|C)\geq
-\sum_{j=1}^{n}p_{j}(\cle;\bro_{A})\,\ln\|\qm_{j}\|_{\infty}
\, , \label{thm5}
\end{equation}
where $\bro_{A}=\tr_{BC}(\bro_{ABC})$ and
\begin{equation}
\qm_{j}=\frac{d^{3}}{n^{3}}\,\sum_{k=1}^{n}
|\phi_{k}\rangle\langle\phi_{k}|\phi_{j}\rangle\langle\phi_{j}|\phi_{k}\rangle\langle\phi_{k}|
=\frac{d^{3}}{n^{3}}\,\bigl((1-c)|\phi_{j}\rangle\langle\phi_{j}|+c\,\sm\bigr)
\, . \nonumber
\end{equation}
As the frame operator reads as $\sm=S\pen_{A}$, we finally have
\begin{equation}
\qm_{j}=S^{-3}\bigl((1-c)|\phi_{j}\rangle\langle\phi_{j}|+Sc\,\pen_{A}\bigr)
\, . \label{qmjs}
\end{equation}
The eigenvalues of this positive operator are equal to
\begin{equation}
\frac{1-c+Sc}{S^{3}}=\frac{d^{3}-2d^{2}+nd}{n^{3}-n^{2}}
 \label{qmjs1}
\end{equation}
with multiplicity 1 and $S^{-2}c$ with multiplicity $d-1$. The
spectral norm of (\ref{qmjs}) is equal to (\ref{qmjs1}), whence
\begin{equation}
\rmh_{1}(X|B)+\rmh_{1}(X|C)\geq
\ln\!\left(
\frac{n^{3}-n^{2}}{d^{3}-2d^{2}+nd}
\right)
 . \label{thm51}
\end{equation}
For an orthonormal basis, the right-hand side of (\ref{thm51}) is
equal to zero as expected. Otherwise, we have a non-trivial bound
for a single ETF-based POVM. It is instructive to compare this bound
with the right-hand side of (\ref{unhh02p}) for $\alpha=1$. The
difference between these right-hand sides is equal to $\ln{S}$. When
the system $A$ is not entangled with others, the result following
from (\ref{thm51}) is weaker than (\ref{unhh02p}). In general, the
two uncertainty relations are independent. It would be interesting
to seek a way to improve (\ref{thm51}). There is a hope to address
this question in a future work.

It is known that entropic uncertainty relations immediately lead to
steering inequalities. Steering is a phenomenon for bipartite
quantum systems that is related to entanglement but is not precisely
the same \cite{wisem07}. For a general discussion of this important
topic, see the review \cite{ucno20} and references therein. Steering
inequalities in terms of the standard entropic functions were
considered in \cite{sbwch13}. The authors of \cite{brun18} focused
on steering inequalities formulated with the use of R\'{e}nyi
entropies. Steering criteria also follow from uncertainty relations
in terms of Tsallis entropies \cite{cug18}. Using quantum designs,
the papers \cite{guhne20,rastdes} derived corresponding uncertainty
relations and their application to steering criteria. Applying the
results of the previous section to quantum steering and comparing
them with similar inequalities taken from the literature is rather
the subject of a separate investigation.

In last years, quantum coherence is intensively studied as an
informational and computational resource. A theoretical framework
for quantifying coherence was developed in \cite{bcp14}. The main
idea is to characterize quantitatively a distinction of the given
density matrix from matrices of the form
\begin{equation}
\bau=\sum_{i=1}^{d} t_{i}\,|b_{i}\rangle\langle{b}_{i}|
\, , \qquad \sum_{i=1}^{d} t_{i}=1
\, . \label{incs}
\end{equation}
Such states are completely incoherent with respect to the
orthonormal basis $\clb=\bigl\{|b_{i}\rangle\bigr\}$. There are
several ways to choose a distinguishability measure that should be
minimized for the given orthonormal basis and state of interest.
Various candidates to quantify coherence and their properties are
reviewed in \cite{sap17}. The quantum relative entropy leads to the
quantity \cite{bcp14}
\begin{equation}
\rmc_{1}(\clb;\bro)=H_{1}(\clb;\bro)-\rmh_{1}(\bro)
\, , \label{c1for}
\end{equation}
which is referred to as the relative entropy of coherence. Since
different computational bases can be preferred depending on the
theoretical or experimental setup, coherence quantifiers are applied
to a variety of bases, including non-orthogonal ones. POVM-based
coherence measures and corresponding incoherent operations were
examined in \cite{bkb2019}. For a rank-one POVM, the question is
naturally resolved with Naimark's extension \cite{rastcm}. In this
case, we merely replace (\ref{c1for}) with
\begin{equation}
\rmc_{1}(\cle;\bro)=H_{1}(\cle;\bro)-\rmh_{1}(\bro)
\, . \label{c1forp}
\end{equation}
Combining this with (\ref{unsh02}), we estimate from
below the relative entropy of coherence with respect to the chosen
ETF. In contrast, the use of other entropic functions does not allow
simple expressions similarly to (\ref{c1for}) and
(\ref{c1forp}).

Finally, we recall the Brukner--Zeilinger approach to quantify an
amount of quantum information \cite{bz1999,bz2002}. Brukner and
Zeilinger proposed an operationally invariant measure of information
in quantum measurements. It was reasoned in \cite{rastpro} that
the difference between two indices of coincidence is a useful measure.
Combining this approach with (\ref{inbr}) gives
\begin{equation}
I(\cle;\bro)-I(\cle;\bro_{*})\leq\frac{d-1}{n^{2}-n}\left[d\,\tr(\bro^{2})-1\right]
 . \label{bzin}
\end{equation}
For a SIC-POVM, the inequality (\ref{bzin}) is saturated so that
both the sides are equal to \cite{rastpro}
\begin{equation}
\frac{\tr(\bro^{2})-\tr(\bro_{*}^{2})}{d(d+1)}
\ . \label{bzin1}
\end{equation}
The numerator of (\ref{bzin1}) is the total information gained in
measurements with the use of $d+1$ MUBs. But the existence of $d+1$
MUBs is proved only when $d$ is a prime power \cite{bz10}. There are
reasons to believe that SIC-POVMs exist for all $d$, and this
conjecture is due to Zauner \cite{zauner11}. Meantime, ETF-based
measurements allow us to estimate the Brukner--Zeilinger measure of
quantum information.

\section{Conclusions}\label{sec7}

We have derived uncertainty relations for quantum measurements
assigned to ETFs. Such frames are interesting in their own rights as
well as due to applications in many disciplines. Complex ETFs
include a class of SIC-POVMs used in quantum information theory. It
is natural to assume that the use of ETFs expands available tools in
building protocols of information processing. The presented results
support this conclusion. Many of them were obtained by a development
of the reasons known for SICs. Thus, ETFs deserve more quantum
applications than they have obtained.

Each of ETFs leads to a POVM-measurement. The inner structure of
ETFs imposes certain restrictions on generated probabilities. In
particular, the index of coincidence and the maximal probability are
bounded from above. Hence, fine-grained and entropic uncertainty
relations with their corollaries were formulated. The elaborated
framework is an extension of the facts found earlier for SICs
\cite{rastmubs}. It is shown that ETF-based measurements allow us to
test quantum correlations. The above results are presented with the
aim to stimulate further use of ETFs in quantum information theory.

\end{document}